\documentclass[conference]{IEEEtran}
\usepackage{helvet}
\IEEEoverridecommandlockouts
\usepackage{cite}
\usepackage{algorithmic}
\usepackage{graphicx}
\usepackage{textcomp}
\usepackage{multirow}
\usepackage{amsfonts}
\usepackage{amsmath}
\usepackage{enumitem}
\usepackage[table,xcdraw]{xcolor}
\usepackage{subfigure}

\usepackage[linesnumbered,ruled,vlined]{algorithm2e}

\def\BibTeX{{\rm B\kern-.05em{\sc i\kern-.025em b}\kern-.08em
		T\kern-.1667em\lower.7ex\hbox{E}\kern-.125emX}}
	
\begin{document}
	
	\title{Autonomous Unknown-Application Filtering and Labeling for DL-based Traffic Classifier Update\\		
	\thanks{\footnotesize This work has been supported in part by the University of Dayton Office for Graduate Academic Affairs through the Graduate Student Summer Fellowship Program.}
}

\author{\IEEEauthorblockN{Jielun Zhang, Fuhao Li, Feng Ye, Hongyu Wu}
\IEEEauthorblockA{{Department of Electrical and Computer Engineering} \\
University of Dayton, Dayton, OH, USA \\\{zhangj46, lif003, fye001, wuh007\}@udayton.edu}
}
	
	\maketitle
	
	\begin{abstract}
		Network traffic classification has been widely studied to fundamentally advance network measurement and management. Machine Learning is one of the effective approaches for network traffic classification. Specifically, Deep Learning (DL) has attracted much attention from the researchers due to its effectiveness even in encrypted network traffic without compromising neither user privacy nor network security. However, most of the existing models are created from closed-world datasets, thus they can only classify those existing classes previously sampled and labeled. In this case, unknown classes cannot be correctly classified. To tackle this issue, an autonomous learning framework is proposed to effectively update DL-based traffic classification models during active operations. The core of the proposed framework consists of a DL-based classifier, a self-learned discriminator, and an autonomous self-labeling model. The discriminator and self-labeling process can generate new dataset during active operations to support classifier update. Evaluation of the proposed framework is performed on an open dataset, i.e., ISCX VPN-nonVPN, and independently collected data packets. The results demonstrate that the proposed autonomous learning framework can filter packets from unknown classes and provide accurate labels. Thus, corresponding DL-based classification models can be updated successfully with the autonomously generated dataset.
	\end{abstract}
	

	\section{Introduction}
	Network packet classification is fundamental for advancing future networks, e.g., in differentiated Quality-of-Service, network intrusion detection (firewall access control), traffic shaping, resource allocation, etc.~\cite{8648137,new1}. However, the huge amount of users and heavy network traffic between multi-source network applications increase the uncertainty of network traffic~\cite{abbas2018mobile,8553650,jzisc}. Such uncertainty further challenges network measurement and management. To combat those issues, data packet classification methods are in need of an advancement. Traditional traffic classifiers that are based on port assignments or the clear text of packet payload become less effective due to dynamic port assignments and encrypted payloads by many applications~\cite{portbased,payloadbased,conventionalbad}. Recently, Machine Learning (ML) algorithms~\cite{SVM,KNN} and Deep Learning (DL) algorithms~\cite{cnntwo,datanet,deeppacket} have been applied to network packet classification. In particular, DL-based traffic classifiers, such as those built around Convolutional Neural Network (CNN), Multilayer Perceptron (MLP), Stacked Autoencoder (SAE), can provide high accuracy (around 95\%) in packet classification even for encrypted applications~\cite{cnntwo,datanet,deeppacket}. Nonetheless, those DL-based traffic classifiers are created with a closed-world assumption, where a classifier is assumed to only identify the classes in a static dataset. In practice, we face an open-world issue, where the number of applications fluctuates and new types of packets are unknown to the classifiers. An existing classifier could be less accurate every time a new application goes on-line. 
	
	To tackle the issue, we propose an autonomous model updating framework to seamlessly update DL-based packet classifiers during active operations. Specially, the proposed framework is capable of (i) filtering packets of unknown applications from active network traffic, (ii) clustering the packets of unknown classes into corresponding discovered classes and assigning labels, (iii) building a new training dataset including both the existing classes and the discovered classes, and (iv) updating the current classifier through transfer learning.
	Our major contributions in this paper can be concluded as follows:
	\begin{itemize}
		\item An autonomous model updating framework is proposed to update DL-based traffic classifiers by generating a new dataset through packet filtering and labeling from unknown classes.
		\item Four distinct DL-based traffic classifiers are implemented to perform accurate traffic classification.
		\item A dataset is built by capturing numerous packets of several popular Internet applications for evaluation and verification of the proposed scheme.
		\item The proposed framework is evaluated in three distinct scenarios where all the built classifiers are used individually.
	\end{itemize}
	
	The rest of the paper is organized as follows. Section~\ref{sec2} summarizes the related work. Section~\ref{sec3} formulates the problem and introduces the preliminaries. Section~\ref{sec4} illustrates the proposed framework. Section~\ref{sec5} demonstrates the evaluation result. Section~\ref{sec6} concludes the work.

	\section{Related Work}\label{sec2}

	\subsection{Traditional Traffic Classification}
	Traditional network traffic classification has been widely studied, such as port-based and payload-based traffic classification approaches~\cite{portbased,payloadbased,conventionalbad}. Port-based approach uses the port assignment of a packet in its TCP/UDP header to match the default port number on the file released by the Internet Assigned Numbers Authority (IANA)~\cite{portbased}.
	Payload-based method, e.g., deep packet inspection~\cite{payloadbased}, inspects the header and the payload for comparing the signatures on the application level.
	Nevertheless, they fail to perform traffic classification due to port translation (i.e., Network Address Port Translation~\cite{conventionalbad}) and encryption of network packets~\cite{payloadbased}.
	
	
	
	\subsection{Machine Learning based Traffic Classification}
	To tackle the issues with the traditional methods, researchers applied both unsupervised ML algorithms (i.e., K-Means, k-nearest neighbors) and supervised ones (e.g., logistic regression, support vector machine)~\cite{mlclassifier,logisticregression,svm1,wangsurvey} to build traffic classifiers, which use packet features such as packet size, inter-arrival time, etc., for classification. Anderson \emph{et al}. in~\cite{logisticregression} proposed to use logistic regression algorithm for learning flow-level features, header information and other features jointly to classify the encrypted traffic as either malware traffic or normal traffic. Saber \emph{et al}. in~\cite{svm1} proposed to combine Principal Component Analysis (PCA)~\cite{PCA} along with the support vector machine for traffic classification relying on only time-based flow features. Their work achieved an average classification accuracy at $94$\% but required a relatively high overhead time to obtain the flow features.
	However, ML-based classifiers usually provide low classification accuracy and they require manual feature selection~\cite{wangsurvey}.
	
	\subsection{Deep Learning based Traffic Classification}
	DL-based traffic classifiers have been widely studied in many researching fields such as computer vision, natural language processing, etc.~\cite{superlong15authoers,neuralnetworkpopular,hongtao,fsnet}. Nonetheless, not until recently did researchers start to build neural networks (e.g., MLP, SAE, CNN) for traffic classification. Li \emph{et al}. in~\cite{hongtao} proposed a byte segment neural network including an attention that extracts features of payload segments individually and outputs classification results with a Softmax-classification layer.
	Moreover, several end-to-end DL-based classification models are proposed~\cite{fsnet,deeppacket,datanet,jzglobecom}. Liu \emph{et al}. in~\cite{fsnet} developed FS-Net based on recurrent neural networks and an autoencoder both traffic classification and packet feature mining. Lotfollahi \emph{et al}. in~\cite{deeppacket} and Wang \emph{et al}. in~\cite{datanet} applied MLP, SAE, and CNN for traffic classification, where the average classification accuracy is above $95$\%. 

	
	
	Apparently, the previous network traffic classification problems were mostly defined based on a closed-world assumption, such that they can only classify packets in a static dataset. In an open-world assumption, these DL-based classifiers need to be updated promptly to provide accurate traffic classification in order to support the network measurement and management.
	
	
	\begin{figure*}[th!] %
		\centering
		\includegraphics[width=7.0 in]{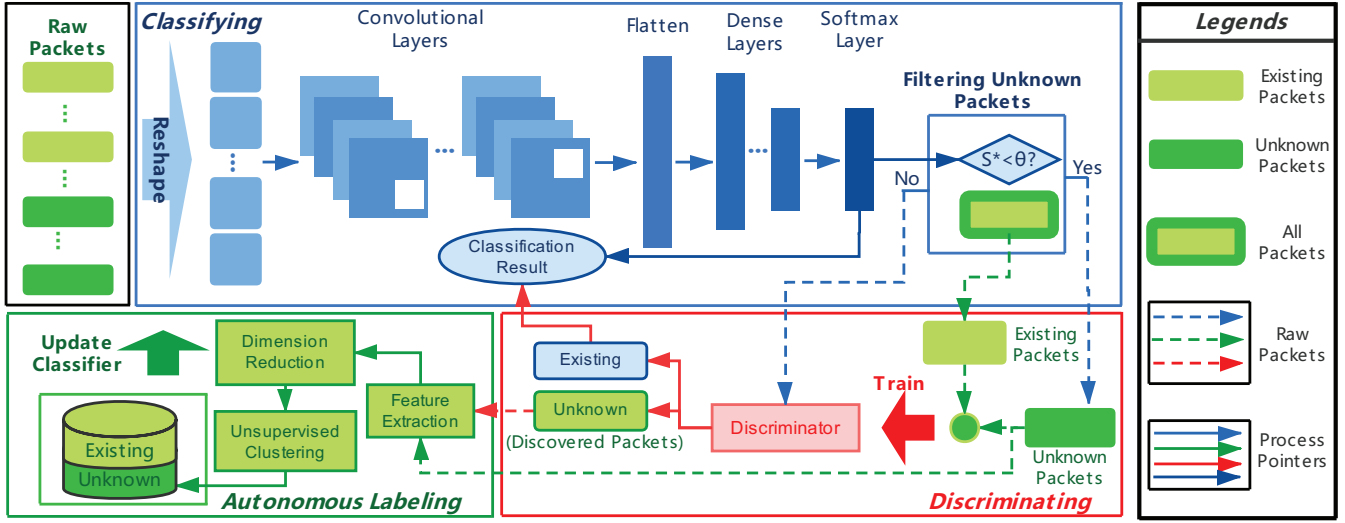}
		\vspace{-2mm}
		\caption{Overview of the proposed autonomous model updating framework.}\label{fig1} 
		\vspace{-4mm}
	\end{figure*}
	
	\section{Problem Formulation}\label{sec3}
	\subsection{Open-world Packet Classification}
	
	
	Let $\mathcal{D} =\{({p}_1,l_1),({p}_2,l_2),...,({p}_n,l_n)\}$ be a set of labeled training dataset that includes a total of $M$ existing application classes. Define ${p}_i$ as the $i$-th instance of application labeled with $l_i \in \mathcal{C} = \{c_1,c_2,...,c_M\}$, and $c_j$ as the corresponding class $j$. A DL-based classification model $F({p}_n) \rightarrow \hat{l}_n$ is created from dataset $\mathcal{D}$ to predict label $\hat{l}_i$ of ${p}_i$ that matches the actual label $l_i$.

	
	Assume that the current DL-based traffic classifier $F^t$ at time $t$ has been created for $M^t$ existing classes of applications i.e., $\mathcal{C}^t = \{c_1,c_2,...,c_{M^t}\}$, using the training dataset $\mathcal{D}^t$. Besides existing classes, we define $\mathcal{U}^t$ as the set of unknown classes that exist in active traffic at time $t$. Thus, the collection of all possible classes of application is $\Omega = \{\mathcal{U}^t \cup \mathcal{C}^t\}$ in the open-world assumption. The classifier is to be updated at a future time $(t+1)$ as $F^{t+1}$ that can discover previously unknown classes of applications $\mathcal{N}^{t+1}$ (defined as \emph{discovered classes}) together with the existing classes. A new dataset $\mathcal{D}^{t+1}$ that comprises both $\mathcal{C}^{t}$ and $\mathcal{N}^{t+1}$ is thus required to update classifier $F^{t+1}$, such that $F^{t+1}({p}_i) \rightarrow \hat{l}_i$, and $\hat{l}_i$ matches the actual label $l_i$ for all $l_i \in \mathcal{C}^{t+1}$. To be specific, three subproblems are formulated to update a DL-based traffic classifier in an open-world classification scenario so that the updated $F^{t+1}$ can correctly identify all $M^{t+1}$ classes of packets in the current active network. The three subproblems are as follows:
	
	\begin{enumerate}
		\item Filtering unknown packets $\mathcal{N}^{t+1}$.
		\item Identify the number of discovered classes.
		\item Update the DL-based traffic classifier from $F^t$ to $F^{t+1}$.
	\end{enumerate}

	\subsection{Preliminaries (Convolutional Neural Network)}
	
	Convolutional Neural Network (CNN) is a popular DL architecture used for image classification~\cite{cnn}. A typical CNN consists of convolutional layers, pooling layers, dense/fully-connected layers and a softmax layer. The convolutional layers perform the feature extraction by convolution kernels. For a data sample $I$ in the format of 2-D matrix, it is processed by convolutional kernels in each convolutional layer as follows:
	\begin{equation}\label{eq2}
	\centering
	\begin{split}
	c[i][j][k] =&q[k] + \sum_{l}\sum_{s=1}^{W}\sum_{t=1}^{H}{w[s][t][l][k]}\\
	&\ast I[(i-1)+s][(j-1)+t][l],\\
	\end{split}
	\end{equation}
	where `$*$' is the convolution operator, $k$ is the order of convolution kennels; $l$ is the channel number of the input; $W$ and $H$ are the width and length of the convolution kernel; $w$ and $q$ are the weights and bias in the corresponding channel. Note that the stride in the illustrated example is set to 1.
	
	The output of a convolutional layer is activated by Rectifier Linear Units (ReLU) for non-linearity. It provides a faster training process and help to avoid gradient vanishing problem compared with other activation functions (i.e., sigmoid and tanh functions)~\cite{relu}. The activation process produced by ReLU is formulated as:
	\begin{equation}\label{fullyconnectedlayerinput}
	x[i][j][k] = \max\left(0,c[i][j][k]\right).
	\end{equation}
	Pooling layers are usually attached to the output of activation functions for dimension reduction, which can speed up the training process. Nonetheless, all the raw packets used in the proposed scheme are relatively small, thus we remove the pooling layers in the proposed DL-based traffic classifiers in order to keep all details in the raw packets. The outputs of the last convolutional layer are flattened and passed to fully connected layers, which are also known as dense layers, where feature maps are produced. Softmax function is widely used at the end of neural networks to map non-normalized output to a probability distribution over classes of prediction~\cite{cnn}. A Softmax layer accepts the output from the last fully-connected layer and provides the final classification result. The fully-connected layer and the Softmax layer are computed as follows:
	\begin{equation}\label{eq5}
	\mathbf{y} = \begin{bmatrix}y_1, y_2, ..., y_N\end{bmatrix} = \left(W^T\cdot \mathbf{v} \right)+\mathbf{b},
	\end{equation}
	\begin{equation}\label{eq6}
	\begin{split}
	\mathbf{s} = \begin{bmatrix}s_1, s_2, .., s_{N}\end{bmatrix} = \frac{\exp(y_n)}{\sum_{i = 1}^{N}{\exp(y_i)}},\\
	\end{split}
	\end{equation}
	where $\mathbf{v}$ is the output of the former dense layer; $\mathbf{y}$ is the output vector of the last dense layer that is connected to the Softmax layer; $s_n$ is the categorical probability for the input to be classified into class $n$, where $s_n \leq 1$ and $\sum{s_n}=1$.

	\section{Autonomous Model Updating Framework}\label{sec4}
	The proposed autonomous model updating framework consists of three stages, including DL-based packet classification, self-learned unknown application discrimination, and autonomous unknown packet self-labeling, as illustrated in Fig.~\ref{fig1}. The classifier is to be deployed at a network gateway for identifying each incoming data packet. It also provides confidence scores (to be detailed in Section~\ref{sectemp}) that are further used in the discriminator. The discriminator is proposed to discriminate the data packets from unknown applications autonomously. Such a process can be triggered either manually or automatically after a period. The application classes of those filtered packets from the discriminator are denoted as \emph{discovered classes}. The discovered classes are further clustered and self-labeled as the new generated dataset. The classification model is eventually updated based on the autonomously generated dataset. The details of each processing stage in the proposed framework are presented as follows.
	
	\subsection{Deep Learning based Traffic Classifier}\label{sectemp}
	Without loss of generality, we assume that there is a DL-based traffic classifier developed based on neural network models, e.g., CNN, MLP, RNN. The classifier first extracts features from an input data packet, and then performs classification by computing the categorical probabilities of the existing classes.
	The classifier is able to accurately identity the existing classes (e.g. $M^t$ classes) based on the training dataset $\mathcal{D}^t$. In an open-world scenario, assume that there are packets from multiple applications of unknown classes in an active network traffic during discrimination.
	 The packets of these unknown application classes are denoted as $\mathbf{p}_{u}$. Apparently, $\mathbf{p}_{u}$ cannot be properly classified by the original classifier.
	
	\begin{figure}[ht!]
		\vspace{-3mm}
		\centering
		\includegraphics[width=3.2in]{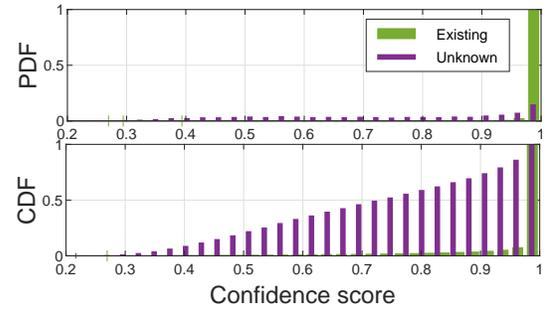}
		\vspace{-3mm}
		\caption{Evaluation result on the classification model.}\label{pdfcdf} 
	\end{figure}
	
 	Let $\mathbf{s}$ be the output of the Softmax layer, and let $s^{\star} = \max(\mathbf{s})$ be the confidence score obtained every time a single packet is processed by the classifier, and $s_c^{\star}$ and $s_u^{\star}$ be the confidence score computed for the packets that belongs to the existing (also known as current) classes and unknown classes correspondingly. Let $\mathbf{P}$ be a group of packets (including $\mathbf{p}_{c}$ and $\mathbf{p}_{u}$) passed through the model during discrimination, and we define $\mathbf{S}_c$ and $\mathbf{S}_u$ as the confidence score sets that contain all the corresponding $s^{\star}_c$ and $s^{\star}_u$ computed from the packets in $\mathbf{P}$. Examples of the probability density function (pdf) and the Cumulative Distribution Function (CDF) of a pair of $\mathbf{S}_c$ and $\mathbf{S}_u$ are illustrated in Fig.~\ref{pdfcdf}. By inspecting the pdf and CDF from extensive experiments, we discovered that most of $s^{\star}_c$ are distributed closely to 1. On the contrary, $s^{\star}_u$ is distributed uniformly since the classifier has been trained only by the samples of existing classes. Such a characterization of the confidence score sets $\mathbf{S}_c$ and $\mathbf{S}_u$ can be applied to filter partial packet samples of unknown classes directly. The pre-filtered packet samples are used to build a set of training data for the discriminator to further cluster and label the unknown classes.

	

	\subsection{Self-learning Discriminator}
	The discriminator is designed as a DL-based binary classifier to distinguish $\mathbf{p}_c$ and $\mathbf{p}_u$.
	Let $\epsilon$ be the classification accuracy of the classifier, and define $\theta$ as the boundary where CDF of $\mathbf{S}_c$ reaches $(1-\epsilon)$. For a packet whose $s^{\star} < \theta$, it will be treated as one the \emph{discovered classes} (formerly unknown classes). Let $\mathbf{P_n}^L$ be a collection of these packets, where $L$ remarks that $\mathbf{P_n}^L$ is a part of the entire set $\mathbf{P_n}$ of all discovered classes whose confidence scores locates on the left side of the boundary. 
	
	\vspace{-2mm}
	\begin{algorithm}[ht!]
		\DontPrintSemicolon
		\SetAlgoLined
		\SetKwInOut{Input}{Input}\SetKwInOut{Output}{Output}
		\Input{$\mathbf{P},\mathcal{D}^t$}
		\Output{$\mathbf{P_n}$}
		\BlankLine
		$\mathbf{P_n^L}$,~$\mathbf{P_n^R}'$,~$\mathbf{P^R}\gets \emptyset$~~~//Initialization\;
		\ForAll{$\mathbf{p}_i \in \mathbf{P}$}{Calculate $\mathbf{s}_i$ based on Eq.~(\ref{eq6})\;Calculate $s_i^{\star} = \max(\mathbf{s}_i)$\; \eIf{$s_i^{\star} \leq \theta$}{$\mathbf{P_n^L} \gets \mathbf{P_n^L} \, || \, p_i$}{$\mathbf{P^R} \gets \mathbf{P^R} \, || \, p_i$}}
		Train discriminator $D({p_i})$ by \{\{$\mathcal{D}^t$,$\mathbf{0}$\},\{$\mathbf{P_n^L}$,$\mathbf{1}$\}\}\;
		\ForAll{${p}_i \in \mathbf{P^R}$}{\If{$D(p_i)==1$}{$\mathbf{P_n^R} \gets \mathbf{P_n^R} \, || \, p_i$}}
		$\mathbf{P_n} \gets \{\mathbf{P_n^L},\mathbf{P_n^R}\}$
		\caption{Filtering packets of unknown classes}\label{alg1}
	\end{algorithm}
\vspace{-2mm}
	
	Once both training samples of $\mathbf{p}_c$ (stored in the database) and  the set $\mathbf{P_n}$ (filtered by the boundary $\theta$) are obtained, the discriminator can be trained and classify the remaining packets in set $\mathbf{P^R}$ whose confidence scores are above the boundary. The packets identified as ones in the unknown classes will be inserted to the set $\mathbf{P_n}^R$. $\mathbf{P_n}^L$ and $\mathbf{P_n}^R$ are merged as the entire set $\mathbf{P_n}$ of all possible discovered classes. Details of the discriminating process are summarized in Alg.~\ref{alg1}.

	
	\subsection{Autonomous Unknown Packet Self-Labeling}  
	
	Although packets of the discovered classes can be filtered by the discriminator, the actual labels of these packets still need to be assigned in order to build a new training set for model update.
	The proposed an autonomous label assigner in an open-world traffic classification scenario performs autonomous labeling in two steps. Firstly, it extracts the feature of packets from the discovered classes in a low dimension. Then, it clusters the extracted features into different groups and label them accordingly. Note that the actual application remains unknown unless the network provider and/or end user are willing to share the information.
	
	
	\subsubsection*{1) Feature Dimension Reduction}~
	\newline
	Feature maps are obtained from a dense layer by passing filtered packets in $\mathbf{P_n}$ through the classifier. They contain the hidden correlations between packets of the unknown classes and the learned packets used to train the classifier previously. Such hidden correlations can thus be used for clustering as the prior knowledge.
	
	Based on the design of classification models, the adoption of feature extraction schemes may be considered to reduce the dimension of the feature for efficient clustering. PCA is adopted in this work due to its computational efficiency. Other dimension reduction schemes can also be used, e.g., Stacked Autoencoder and Reconstruction Independent Component Analysis. Let $\mathbb{Y}\in \mathbb{R}^{W\times V}$ be the combination of the feature maps obtained from $V$ packets in $\mathbf{P_n}$, such that:
	\begin{equation}
	\begin{split}
	\mathbb{Y} &= \begin{bmatrix}
	{\mathbf{y}_1}, {\mathbf{y}_2}, \dots, {\mathbf{y}_V}
	\end{bmatrix}=\begin{bmatrix}
	y_{1,1} & y_{1,2} & \dots  & y_{1,V} \\
	y_{2,1} & y_{2,2} & \dots  & y_{2,V} \\
	\vdots & \vdots & \ddots & \vdots \\
	y_{W,1} & y_{W,2} & \dots  & y_{W,V}
	\end{bmatrix} 
	\end{split},
	\end{equation}
	where $\mathbf{y}_v$ is the feature vector of $v$-th packet in $\mathbf{P_n}$ that contains a total of $W$ feature values, and $W$ is the number of neurons in the dense layer. Mapminmax Normalization is performed as follows to normalize $y_{w,v}$, s.t.,
	\begin{equation}
	y_{w,v} = \frac{y_{w,v} - \min(\mathbf{y_v})}{\max(\mathbf{y_v}) - \min{(\mathbf{y_v})}}.
	\end{equation}
	The result of the normalization would be $0$ if $\mathbf{y_v}$ is a zero vector. The covariance matrix is defined as follows:
	\begin{equation}
	G = \frac{1}{W-1}\sum_{i=1}^{W}{\left(\mathbf{y_v} - \mu_v\right)\left(\mathbf{y_v} - \mu_v\right)^\text{T}},
	\end{equation}
	where $u_v$ is the mean of $\mathbf{y_v}$. We then compute the eigenvector $U = [U_1, U_2, \dots, U_W]$ of $G$ s.t. $(\lambda I - G)U = 0$, where $\lambda=[\lambda_1,\ldots,\lambda_W]$ are eigenvalues. Note that $\lambda_W$ are rearranged in descending order, i.e., $\lambda_1\ge \lambda_2 \ge \ldots \ge \lambda_W$. Let $H$ denote the extracted principal components as follows:
	\begin{equation}
	H = U^\text{T}\mathbb{Y}.
	\end{equation}
	The first $q$ columns of $H$ can be chosen as the representation of the principal information for the feature set $\mathbb{Y}$, where $q$ defines the dimension of the extracted principal components.
	
	\subsubsection*{2) Autonomous Clustering}~
	\newline
	The extracted features are then clustered into a few groups that represent high similarity among the other packets in the same group. For simplicity, K-mean~\cite{kmeans} is applied in this work for demonstration. In theory, other clustering algorithms can be applied in the process. Assume that the centroids (centers of the clusters) of the $i$-th discovered class $\mathbf{N}_i$ after the clustering is marked as $\mathbf{O}^i$, which is calculated as follows:
	\begin{equation}
	\mathbf{O}^i = \frac{1}{|\mathbf{N}_i|} \cdot \sum_{\mathbf{o}_j \in \mathbf{N}_i}{\mathbf{o}_j},
	\end{equation}
	where $|\mathbf{N}_i|$ is the total number of samples in class $\mathbf{N}_i$. The similarity between principal components of the network traffic from two applications can be measured by their Euclidean Distance, calculated as follows:
	\begin{equation}
	d(\mathbf{o}_a,\mathbf{o}_b) = \sqrt{{(\mathbf{o}_a - \mathbf{o}_b)^2}}.
	\end{equation}
	A smaller $d$ represents a closer relationship, in contrast, a larger $d$ value represents a lower similarity between them. The converged centroids will be used to cluster packets in $\mathbf{P_n}$. The objective function for the clustering model is defined as follows:
	\begin{equation}
	\min~\sum_{\mathbf{o}_j \in \mathbf{N}_i}{\vert\mathbf{o}_j -\mathbf{O}^i\vert}^2.
	\end{equation}
	For autonomous clustering, Bayesian information criterion (BIC)~\cite{bic} is applied to find the optimal number of group, denoted as $K$. The BIC in our proposed clustering problem is calculated as follows:
	\begin{equation}\label{eq11}
	\text{BIC} = V\cdot \ln\left(\frac{R}{V}\right) +k\cdot \ln\left(V\right),
	\end{equation}
	\begin{equation}\label{eq10}
	R = \sum_{i = 1}^{k}\sum_{\mathbf{o}_j\in \mathbf{N}_i}\sqrt{(\mathbf{o}_j-\mathbf{O}^i)^2},
	\end{equation}
	where $V$ is the number of samples in $\mathbf{P_n}$ to be clustered; $k$ is the index of clusters for enumeration; and $R$ is the sum of root squared errors.
	Let $K_{\max}$ be the upper bound of the groups, such that $K_{\max} \le V$. $K_{\max}$ can be defined based on the network environment.
	In each test of index $k$, a BIC value is calculated for clustering model evaluation. After finishing all enumeration of $k$ from $1$ to $K_{\max}$, the clustering model provides the most BIC decreasing is considered as the optimal one and the number of clusters included in it is the optimal cluster number. The overall autonomous clustering algorithm is summarized in Alg.~\ref{alg2}, where $\mathbf{o}$ is the set of all extracted features; $\mathcal{M}$ is the clustering model with converged cluster centroids; $\mathcal{M}^*$ is the optimal clustering model; $l_k \in \mathbf{l_n}$ are the new assigned labels to the packets of discovered classes in $\mathbf{P_n}$.
	
	\begin{algorithm}[ht!]\label{alg2}
		\KwData{$\mathbf{o}$, $K_{\max}$}
		\KwResult{$\mathcal{M}^*$,$\mathbf{l_n}$}
		initialization\;
		\textbf{For} $k = 1:K_{\max}$\\
		Randomly choose $k$ objects from $\mathbf{O}$ as the initial centers of clusters of the new classes\;
		\textbf{Repeat:}\\
		1) Assign or reassign each $\mathbf{o}_i$ to the cluster to which the $\mathbf{o}_i$ is the most similar, based on the mean value of all $\mathbf{o}_i$ in the cluster\;
		2) Update the cluster centroids by calculating the mean value of the $\mathbf{o}_i$ for each cluster\;
		\textbf{Until} Cluster centroids convergence, or reach the assigned maximum iteration time\;
		Save the current model $\mathcal{M}_k$\;
		Compute BIC$_k$ based on Eq.~(\ref{eq10}), (\ref{eq11}) for $\mathcal{M}_k$\;
		Calculate $\Delta \text{BIC} = \text{BIC}_k - \text{BIC}_{k-1}$ for $k>1$\;
		\textbf{EndFor}\\
		Find $\mathcal{M}_K$ obtains $\max(\Delta \text{BIC})$\;
		$\mathcal{M}^* \gets \mathcal{M}_K$\;
		Assign label $l_{k}$ to all packets clustered in the $k$-th unknown class based on $\mathcal{M}^*$\; 
		Store $l_k$ in the label set $\mathbf{l_n}$ with respect to the packets in $\mathbf{P_n}$\;
		\caption{Autonomous application clustering}
	\end{algorithm}

	
	
	
	\subsection{New Dataset Generation and Model Update}
	The new dataset $\mathcal{D}^{t+1}$ is composed of both the old dataset $\mathcal{D}^t$ and the newly discovered ones (denoted as \{$\mathbf{P_n}$,~$\mathbf{l_n}$\}), such that $\mathcal{D}^{t+1} \gets \mathcal{D}^t$  $\, || \,$  \{$\mathbf{P_n}$,~$\mathbf{l_n}$\}. To update the classifier from current state $F^t$ to $F^{t+1}$, transfer learning scheme is adopted since it migrates parameters of the current model to boost the training process. The detailed transfer learning process is presented in Alg.~\ref{alg3}.
	
	\vspace{-2mm}
	\begin{algorithm}[ht!]
		\DontPrintSemicolon
		\SetAlgoLined
		\SetKwInOut{Input}{Input}\SetKwInOut{Output}{Output}
		\Input{${F}^{t}$, $\mathcal{D}^{t+1}$}
		\Output{$F^{t+1}$}
		\BlankLine
		
		\While{the model updating interval lapses}{
			Load new dataset $\mathcal{D}^{t+1}$ in the database\;
			Load layers and weights in $F^t$\;
			Resize the last dense layer from $M^t$ to $M^{t+1}$\;
			Resize Softmax layer from length $M^t$ to $M^{t+1}$\;
			Save the modified layers for ${F}^{t+1}$\;
			Train $F^{t+1}$ with the new dataset $\mathcal{D}^{t+1}$\;
		}
		\caption{Classifier Update}\label{alg3}
	\end{algorithm}
	\vspace{-2mm}

	\section{Evaluation}\label{sec5}
	
	
	\subsection{Dataset for Evaluation}
	Part of the evaulation dataset is selected from ``ISCX VPN-nonVPN dataset'' (ISCXVPN2016)~\cite{ISCX}. A total of $206,688$ packets, including Skype, Youtube, Vimeo, etc.~\cite{ISCX,datanet} are extracted from the dataset. Those applications are encrypted by different security protocols, e.g., HTTPS, SSL, SSH, etc. To further evaluate the proposed model updating scheme, we also collected a dataset from real-life network applications with encrypted packets. To ensure rich diversity and quantity, we capture a total of $492,721$ packets from $5$ distinct applications, including Google Map, Speedtest by Ookla, Tencent QQ, Discord and DOTA2. 
	Details of the dataset is summarized in Table~\ref{tb2}. Note that the first 24 bytes of each packet are removed to focus on encrypted payload only.
	\begin{table}[th!]
		\vspace{-2mm}
		\caption{Summary of the dataset used for evaluation.}\label{tb2}
		\begin{tabular}{|c|c|c|c|}
			\hline
			\textbf{Application}          & \textbf{\begin{tabular}[c]{@{}c@{}}Total \#  \\samples\end{tabular}} & \textbf{Application} & \textbf{\begin{tabular}[c]{@{}c@{}}Total \#  \\samples\end{tabular}} \\ \hline
			\textit{Google Map$^*$}           & $54,114$                                                                & Netflix              & $51,932$                                                                \\ \hline
			\textit{Speedtest (upload)$^*$}   & $112,354$                                                               & SCP (download)       & $15,390$                                                                \\ \hline
			\textit{Speedtest (download)$^*$} & $39,302$                                                                & SFTP (download)      & $4,729$                                                                 \\ \hline
			\textit{Discord$^*$}              & $20,032$                                                                & Skype file           & $4,607$                                                                 \\ \hline
			\textit{Tencent QQ (voice)$^*$}   & $143,370$                                                               & TorTwitter           & $14,654$                                                                \\ \hline
			\textit{DOTA2$^*$}                & $123,549$                                                               & Vimeo                & $18,755$                                                                \\ \hline
			Email clients                 & $4,417$                                                                 & VOIPbuster           & $35,469$                                                                \\ \hline
			Facebook chat                 & $5,527$                                                                 & Youtube              & $12,738$                                                                \\ \hline
		\end{tabular}
	\vspace{+1mm}
		\newline
		\textbf{Note:} Applications marked with $*$ in italics are new source applications for packet collection. The others are sampled from ISCX VPN-nonVPN dataset.
	\end{table}
\vspace{-4mm}

	\subsection{Design of Classification Model}
	
\begin{figure}[ht!] 
		\centering
		\includegraphics[width=3.in]{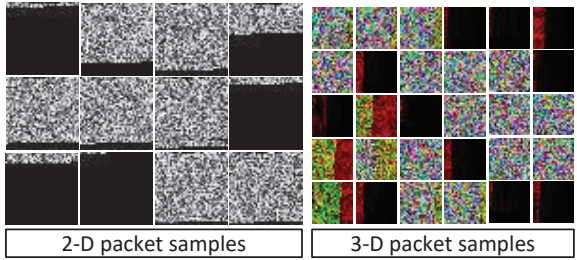}
		\vspace*{-3mm}
		\caption{Examples of input packets to 2-D CNN and 3-D CNN classifiers.}\label{image} 
	\end{figure}

	The input of our proposed 1-D classifier is a packet vector with a dimension of $1\times 1456$ bytes. 
	In our designed 2-D classifier, packet vectors are reshaped to $39 \times 39$ and can be visualized as gray images (see examples on the left-hand side of Fig.~\ref{image}). In our designed 3-D classifier, packet vectors are converted in to a 3D tensor with a size of  $22\times 22\times 3$, which can be visualized as $24$-bit RGB images (see examples on the right-hand side of Fig.~\ref{image}). Detailed specifications of converted packets and the built classification models are summarized in the Table~\ref{tb1}.
	
	\begin{table}[ht!]
			\vspace{-2mm}
		\caption{Specifications of the proposed deep learning based network traffic classification models.}\label{tb1}
		\vspace*{-2mm}
		\centering
		\begin{tabular}{|c|c|c|c|c|}
			\hline
			\textbf{Classifier type}   & \textit{MLP}  & \textit{1-D CNN}  & \textit{2-D CNN}& \textit{3-D CNN}\\ \hline
			\textbf{Input size}& 1$\times$1456 & 1$\times$1456& 39$\times$39 & 22$\times$22$\times$3  \\ \hline
			\textbf{Input length} & 1456          & 1456   & 1521  & 1452 \\ \hline
			\textbf{\begin{tabular}[c]{@{}c@{}}Convolutional\\kernel size\end{tabular}}  & - & \begin{tabular}[c]{@{}c@{}}(1$\times$9)$\times$16\\ (1$\times$9)$\times$16\end{tabular} & \begin{tabular}[c]{@{}c@{}}(3$\times$3)$\times$64\\ (3$\times$3)$\times$32\end{tabular} &  \begin{tabular}[c]{@{}c@{}}(3$\times$3$\times$3)$\times$64\\ (3$\times$3$\times$3)$\times$32\end{tabular} \\ \hline
			\textbf{\begin{tabular}[c]{@{}c@{}}Activation\\ function\end{tabular}}   & ReLU    &ReLU & ReLU &ReLU\\ \hline
			\textbf{\begin{tabular}[c]{@{}c@{}}Sizes of \\dense layers\end{tabular}} & \begin{tabular}[c]{@{}c@{}}768\\ 128\\9\end{tabular} & \begin{tabular}[c]{@{}c@{}}128\\ 9\end{tabular} & \begin{tabular}[c]{@{}c@{}}128\\ 9\end{tabular} &  \begin{tabular}[c]{@{}c@{}}128\\ 9\end{tabular} \\ \hline
			\textbf{Softmax layer}  & Yes& Yes& Yes & Yes  \\ \hline
		\end{tabular}
	\end{table}
	
	
	\subsection{Experiment Settings}
	
	\subsubsection{Experiment Environment}
	
	The evaluation and simulation of the proposed schemes are conducted on a workstation with an Intel$\textsuperscript{\textregistered}$ Xeon$\textsuperscript{\textregistered}$ CPU E5-2630 v3 @ 2.40GHz, $32.0$ GB RAM @ $2133$~MHz, a $480$~GB SSD and an NIVIDIA GeForce GTX $1080$ Ti. Matlab 2019a running in Windows 10 Enterprise is used for the scheme implementation. We also use another graphic processing unit, i.e., a docked NIVIDIA GeForce GTX $1080$ connected to an ultrabook through Thunderbolt 3, to evaluate the processing speed of the classifiers.

	\subsubsection{Evaluation Metrics}
		
	Recall, Precision and F measurement score are applied to evaluate both the proposed clustering scheme and the updated classifiers. We define a true positive (TP) decision assigns two similar packets to the same cluster, a true negative (TN) decision assigns two dissimilar packets to different clusters. A classifier may have two types of erroneous outputs. One is the false positive (FP) decision, which assigns two dissimilar packets to the same cluster. The other one is false negative (FN) decision, which assigns two similar packets to different clusters. The evaluation metrics are formulated as follows:
	\begin{equation}
	R = \frac{\text{TP}}{\text{TP}+\text{FN}},~~P = \frac{\text{TP}}{\text{TP}+\text{FP}},~~F_{\beta} = \frac{({\beta}^2+1)P R}{{\beta}^2 P+R},
	\end{equation}
	where $\beta > 1$ can be used as the penalty factor to provide more weight to recall, and we choose $\beta = 1$ in this paper. 
	
	 To evaluate the clustering performance, let TP be the number of true positive instances properly classified as \textbf{X}; TN be the number of true negative instances properly classified as \textbf{not X}; FP be the number of false positive instances classified as \textbf{X} incorrectly; and FN be the number of false negative instances classified as \textbf{not X} incorrectly. The Rand Index (RI)~\cite{ri} is used, which is defined as follows:
	\begin{equation}
	\text{RI} = \frac{\text{TP}+\text{TN}}{\text{TP}+\text{FP}+\text{FN}+\text{TN}}.
	\end{equation}The RI provides equal weight to FP and FN instances. To penalize FN instances over FP instances for focusing on clustering similar packets to the same cluster as much as possible, F measure can be applied instead.

\subsection{Evaluation Results}
To demonstrate the effectiveness and robustness of our proposed autonomous classifier updating scheme, evaluations are conducted in three distinct scenarios, as detailed in Table~\ref{tb3}. 

\begin{table}[ht!]		
	\vspace{-2mm}
	\caption{Evaluation Scenarios.}\label{tb3}
	\begin{tabular}{|c|c|c|c|}
		\hline
		& \multicolumn{3}{c|}{\textbf{Scenario}}                                                                                                                                                                                                                                                                                                                                                                                                                                                                                        \\ \hline
		& \textbf{A}                                                                                                                                                                 & \textbf{B}                                                                                                                                                              & \textbf{C}                                                                                                                                                             \\ \hline
		\textbf{\begin{tabular}[c]{@{}c@{}}Existing\\ classes\end{tabular}} & \begin{tabular}[c]{@{}c@{}}Email clients, \\ Youtube,\\ Vimeo, \\ Skype file,\\ SFTP (down), \\ TorTwitter,\\ Facebook chat, \\ VOIPbuster,\\ SCP (down).\end{tabular} & \begin{tabular}[c]{@{}c@{}}Skype file, \\ Facebook chat,\\ VOIPbuster, \\ Youtube,\\ DOTA2*, \\ Email clients,\\ Vimeo, \\ SFTP (down),\\ STest (up)*.\end{tabular} & \begin{tabular}[c]{@{}c@{}}Youtube, \\ Facebook chat,\\ Email clients, \\ Skype file,\\ Vimeo, \\ SFTP (down),\\ TorTwitter, \\ VOIPbuster,\\ SCP (down).\end{tabular} \\ \hline
		\textbf{\begin{tabular}[c]{@{}c@{}}Unknown\\ classes\end{tabular}}  & \begin{tabular}[c]{@{}c@{}}Discord*, \\ Google Map*.\end{tabular}                                                                                                          & \begin{tabular}[c]{@{}c@{}}TorTwitter, \\ SCP (down).\end{tabular}                                                                                                      & \begin{tabular}[c]{@{}c@{}}Netflix, \\ STest (down)*, \\ QQ (voice)*.\end{tabular}                                                                         \\ \hline			
	\end{tabular}     
	\vspace{+1mm}                                                                  
	\newline
	\textbf{Note:} packets of the application marked with $*$ are selected from our captured dataset; STest stands for Speedtest; `up' and `down' represent upload and download respectively.
\end{table}

\begin{table*}[ht!]
	\centering
	\caption{Performance evaluation results of the classifiers.}\label{tb4}
	\begin{tabular}{|c|c|c|c|c|c|c|c|c|c|c|c|c|c|c|c|c|}
		\hline
		\multicolumn{2}{|c|}{\multirow{2}{*}{\textbf{}}} & \multicolumn{3}{c|}{\textbf{Recall}} & \multicolumn{3}{c|}{\textbf{Precision}} & \multicolumn{3}{c|}{\textbf{F1 score}} & \multicolumn{3}{c|}{\textbf{Speed (packets/ms)}} & \multicolumn{3}{c|}{\textbf{Bandwidth (Mbps)}} \\ \cline{3-17} 
		\multicolumn{2}{|c|}{}                           & Max.       & Avg.       & Min.       & Max.        & Avg.        & Min.        & Max.        & Avg.        & Min.       & GPU1           & GPU2           & CPU            & GPU1           & GPU2          & CPU           \\ \hline
		\multicolumn{17}{|c|}{\textbf{Scenario A}}                                                                                                                                                                                                                                     \\ \hline
		\multirow{2}{*}{\textbf{MLP}}       & Original   & $95.0$     & $94.5$     & $94.1$     & $95.0$      & $94.5$      & $94.1$      & $95.0$      & $94.5$      & $94.1$     & $19.7$         & $21.3$         & $12.4$         & $229.3$        & $236.8$       & $144.6$       \\ \cline{2-17} 
		& Updated    & $91.1$     & $90.7$     & $90.0$     & $91.5$      & $91.1$      & $90.4$      & $91.3$      & $90.9$      & $90.2$     & $21.2$         & $20.4$         & $11.6$         & $247.3$        & $238.1$       & $136.2$       \\ \hline
		\multirow{2}{*}{\textbf{1D-CNN}}    & Original   & $97.7$     & $97.1$     & $96.7$     & $97.6$      & $97.1$      & $96.7$      & $97.6$      & $97.1$      & $96.8$     & $6.1$          & $8.7$          & $0.4$          & $70.7$         & $101.8$       & $4.7$         \\ \cline{2-17} 
		& Updated    & $94.5$     & $93.9$     & $93.2$     & $95.2$      & $94.7$      & $94.2$      & $94.8$      & $94.3$      & $93.7$     & $5.9$          & $8.5$          & $0.4$          & $68.9$         & $99.9$        & $4.6$         \\ \hline
		\multirow{2}{*}{\textbf{2D-CNN}}    & Original   & $97.7$     & $97.0$     & $96.6$     & $97.7$      & $97.1$      & $96.6$      & $97.7$      & $97.0$      & $96.6$     & $7.3$          & $10.3$         & $0.4$          & $89.2$         & $125.6$       & $5.3$         \\ \cline{2-17} 
		& Updated    & $96.7$     & $96.3$     & $95.7$     & $96.9$      & $96.5$      & $95.9$      & $96.8$      & $96.4$      & $95.8$     & $8.1$          & $10.4$         & $0.4$          & $98.4$         & $126.7$       & $5.3$         \\ \hline
		\multirow{2}{*}{\textbf{3D-CNN}}    & Original   & $97.4$     & $97.1$     & $96.7$     & $97.4$      & $97.1$      & $96.7$      & $97.4$      & $97.1$      & $96.7$     & $13.6$         & $15.1$         & $0.9$          & $158.1$        & $175.7$       & $10.9$        \\ \cline{2-17} 
		& Updated    & $96.5$     & $96.0$     & $95.6$     & $96.6$      & $96.1$      & $95.6$      & $96.5$      & $96.1$      & $95.5$     & $14.5$         & $15.6$         & $0.9$          & $168.1$        & $181.4$       & $10.9$        \\ \hline
		\multicolumn{17}{|c|}{\textbf{Scenario B}}                                                                                                                                                                                                                                     \\ \hline
		\multirow{2}{*}{\textbf{MLP}}       & Original   & $93.2$     & $92.3$     & $91.7$     & $93.2$      & $92.3$      & $91.7$      & $93.2$      & $92.3$      & $91.6$     & $20.3$         & $21.3$         & $12.5$         & $236.6$        & $249.1$       & $145.2$       \\ \cline{2-17} 
		& Updated    & $92.0$     & $91.3$     & $90.6$     & $92.1$      & $91.4$      & $90.7$      & $92.0$      & $91.3$      & $90.6$     & $21.7$         & $20.4$         & $12.3$         & $252.6$        & $237.9$       & $142.8$       \\ \hline
		\multirow{2}{*}{\textbf{1D-CNN}}    & Original   & $97.7$     & $97.4$     & $97.1$     & $97.7$      & $97.4$      & $97.0$      & $97.8$      & $97.4$      & $97.0$     & $5.3$          & $8.7$          & $0.4$          & $61.2$         & $101.8$       & $4.8$         \\ \cline{2-17} 
		& Updated    & $96.5$     & $96.0$     & $95.6$     & $96.5$      & $96.1$      & $95.7$      & $96.5$      & $96.0$      & $95.7$     & $6.5$          & $8.7$          & $0.4$          & $75.4$         & $101.3$       & $4.7$         \\ \hline
		\multirow{2}{*}{\textbf{2D-CNN}}    & Original   & $98.2$     & $97.9$     & $97.4$     & $98.2$      & $97.9$      & $97.4$      & $98.2$      & $97.9$      & $97.4$     & $8.7$          & $10.2$         & $0.4$          & $105.9$        & $125.1$       & $5.3$         \\ \cline{2-17} 
		& Updated    & $96.6$     & $96.2$     & $95.7$     & $96.6$      & $96.2$      & $95.7$      & $96.6$      & $96.2$      & $95.7$     & $8.9$          & $10.0$         & $0.4$          & $108.1$        & $121.5$       & $5.0$         \\ \hline
		\multirow{2}{*}{\textbf{3D-CNN}}    & Original   & $97.6$     & $97.3$     & $96.8$     & $97.6$      & $97.3$      & $96.8$      & $97.6$      & $97.3$      & $96.8$     & $12.2$         & $15.2$         & $1.0$          & $142.1$        & $176.1$       & $11.0$        \\ \cline{2-17} 
		& Updated    & $96.4$     & $96.0$     & $95.6$     & $96.5$      & $96.0$      & $95.6$      & $96.4$      & $96.0$      & $95.6$     & $14.1$         & $15.3$         & $0.9$          & $163.9$        & $178.8$       & $10.5$        \\ \hline
		\multicolumn{17}{|c|}{\textbf{Scenario C}}                                                                                                                                                                                                                                     \\ \hline
		\multirow{2}{*}{\textbf{MLP}}       & Original   & $92.4$     & $91.8$     & $91.1$     & $92.4$      & $91.8$      & $91.1$      & $92.4$      & $91.8$      & $91.1$     & $20.8$         & $20.6$         & $12.5$         & $242.1$        & $240.1$       & $146.1$       \\ \cline{2-17} 
		& Updated    & $92.5$     & $91.8$     & $91.1$     & $92.6$      & $92.0$      & $91.1$      & $92.5$      & $91.9$      & $91.2$     & $20.6$         & $20.9$         & $11.7$         & $239.4$        & $243.0$       & $136.6$       \\ \hline
		\multirow{2}{*}{\textbf{1D-CNN}}    & Original   & $98.0$     & $97.6$     & $97.2$     & $98.0$      & $97.6$      & $97.2$      & $98.0$      & $97.6$      & $97.2$     & $6.2$          & $8.7$          & $0.4$          & $72.1$         & $101.6$       & $4.7$         \\ \cline{2-17} 
		& Updated    & $97.8$     & $97.5$     & $97.0$     & $97.8$      & $97.5$      & $97.0$      & $97.8$      & $97.5$      & $97.0$     & $6.3$          & $8.6$          & $0.4$          & $72.9$         & $100.0$       & $4.5$         \\ \hline
		\multirow{2}{*}{\textbf{2D-CNN}}    & Original   & $98.2$     & $97.9$     & $97.5$     & $98.2$      & $97.9$      & $97.6$      & $98.2$      & $97.9$      & $97.5$     & $7.7$          & $10.3$         & $0.4$          & $93.5$         & $125.9$       & $5.3$         \\ \cline{2-17} 
		& Updated    & $98.1$     & $97.8$     & $97.3$     & $98.1$      & $97.8$      & $97.3$      & $98.1$      & $97.8$      & $97.3$     & $7.6$          & $10.0$         & $0.4$          & $93.0$         & $122.1$       & $5.2$         \\ \hline
		\multirow{2}{*}{\textbf{3D-CNN}}    & Original   & $97.8$     & $97.4$     & $97.0$     & $97.8$      & $97.4$      & $97.0$      & $97.8$      & $97.4$      & $97.0$     & $13.7$         & $15.0$         & $0.9$          & $159.5$        & $174.4$       & $10.9$        \\ \cline{2-17} 
		& Updated    & $97.4$     & $97.1$     & $96.7$     & $97.4$      & $97.1$      & $96.7$      & $97.4$      & $97.1$      & $96.7$     & $13.2$         & $14.9$         & $0.9$          & $153.8$        & $173.4$       & $10.8$        \\ \hline
	\end{tabular}
\end{table*}


In each scenario, we compose five data portions $\mathcal{P}_1$, $\mathcal{P}_2$, $\mathcal{P}_3$, $\mathcal{P}_4$, and $\mathcal{P}_5$ by selecting packets randomly in the combined dataset which including both ISCX dataset and the newly collected dataset. 
Data portion $\mathcal{P}_1$ is the training dataset that consists of $2,500$ packets randomly chosen from each existing application class. Data portion $\mathcal{P}_3$ is the validation dataset that consists of $1,200$ packets from each existing application. Data portions $\mathcal{P}_2$ and $\mathcal{P}_4$ consist of $500$ packets from each existing application and $4,500$ packets from each unknown application. Data portion $\mathcal{P}_4$ represents the active unknown packets. It is used along with data portion $\mathcal{P}_2$ as well as partial data in portion $\mathcal{P}_1$ ($500$ packets from each class) to simulate the active network traffic. The combination of the data portions allow the discriminator to learn the relationship between existing classes and unknown classes autonomously. Portion $\mathcal{P}_5$ is comprised of $1,200$ packets from unknown classes. It is combined with portion $\mathcal{P}_4$ and portion $\mathcal{P}_3$ to test the updated classifier.

For better illustration, we choose to compare the updating performance of the proposed scheme among MLP, 1D-CNN, 2D-CNN and 3D-CNN based classifiers in all proposed scenarios. The summary of classification performance is given in Table~\ref{tb4}. We use two different GPUs for the evaluation, namely, GPU1 (NVIDIA GeForce GTX 1080) and GPU2 (NVIDIA GeForce GTX 1080 Ti), as well as the CPU (Intel$\textsuperscript{\textregistered}$ Xeon$\textsuperscript{\textregistered}$ CPU E5-2630 v3 @ 2.40GHz) to evaluate the computational efficiency. 50 rounds of tests are conducted for both original and updated classifiers. 500 packets of each application are randomly chosen in each round from the testing dataset. It can be observed that all CNN based classifiers outperform the MLP one. Nonetheless, the MLP leads in computational efficiency, followed by 3D-CNN. The lightweight network structure of the MLP-based classifier allows it to support a higher bandwidth. It can be concluded from Table~\ref{tb4} that the 3D-CNN based classifier has the best overall performance because of its high classification accuracy and computational efficiency.

\begin{figure}[th!]
\centering
\includegraphics[width=3.4in]{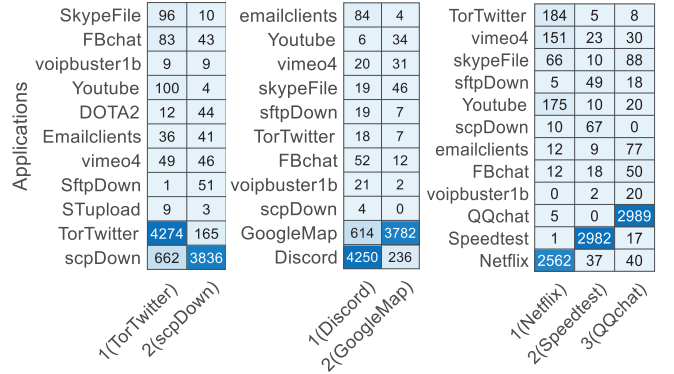}
\vspace{-2mm}
\caption{Clustering results in the testing scenarios (3D-CNN).}\label{fig:awesome_image1}
\vspace{-2mm}
\end{figure}
\begin{table}[th!]
\centering
\caption{Autonomous labeling performance in 3D-CNN classifier.}\label{tb6}
\begin{tabular}{cclclcll}
	\hline
	\multicolumn{1}{|c|}{\textbf{3D-CNN}}                                                            & \multicolumn{2}{c|}{\textbf{Recall}} & \multicolumn{2}{c|}{\textbf{Precision}} & \multicolumn{2}{c|}{\textbf{F1 score}} & \multicolumn{1}{l|}{\textbf{Rand Index}} \\ \hline
	\multicolumn{1}{|c|}{\textbf{\begin{tabular}[c]{@{}c@{}}Scenario A\end{tabular}}} & \multicolumn{2}{c|}{83.1}            & \multicolumn{2}{c|}{75.6}               & \multicolumn{2}{c|}{79.2}              & \multicolumn{1}{c|}{79.9}                \\ \hline
	\multicolumn{1}{|c|}{\textbf{\begin{tabular}[c]{@{}c@{}}Scenario B\end{tabular}}} & \multicolumn{2}{c|}{83.7}            & \multicolumn{2}{c|}{72.0}               & \multicolumn{2}{c|}{77.4}              & \multicolumn{1}{c|}{78.7}                \\ \hline
	\multicolumn{1}{|c|}{\textbf{\begin{tabular}[c]{@{}c@{}}Scenario C\end{tabular}}} & \multicolumn{2}{c|}{94.4}            & \multicolumn{2}{c|}{83.4}               & \multicolumn{2}{c|}{88.5}              & \multicolumn{1}{c|}{92.8}                \\ \hline                   
\end{tabular}
\end{table}

\begin{figure*}[ht!] 
\centering
\includegraphics[width=7in]{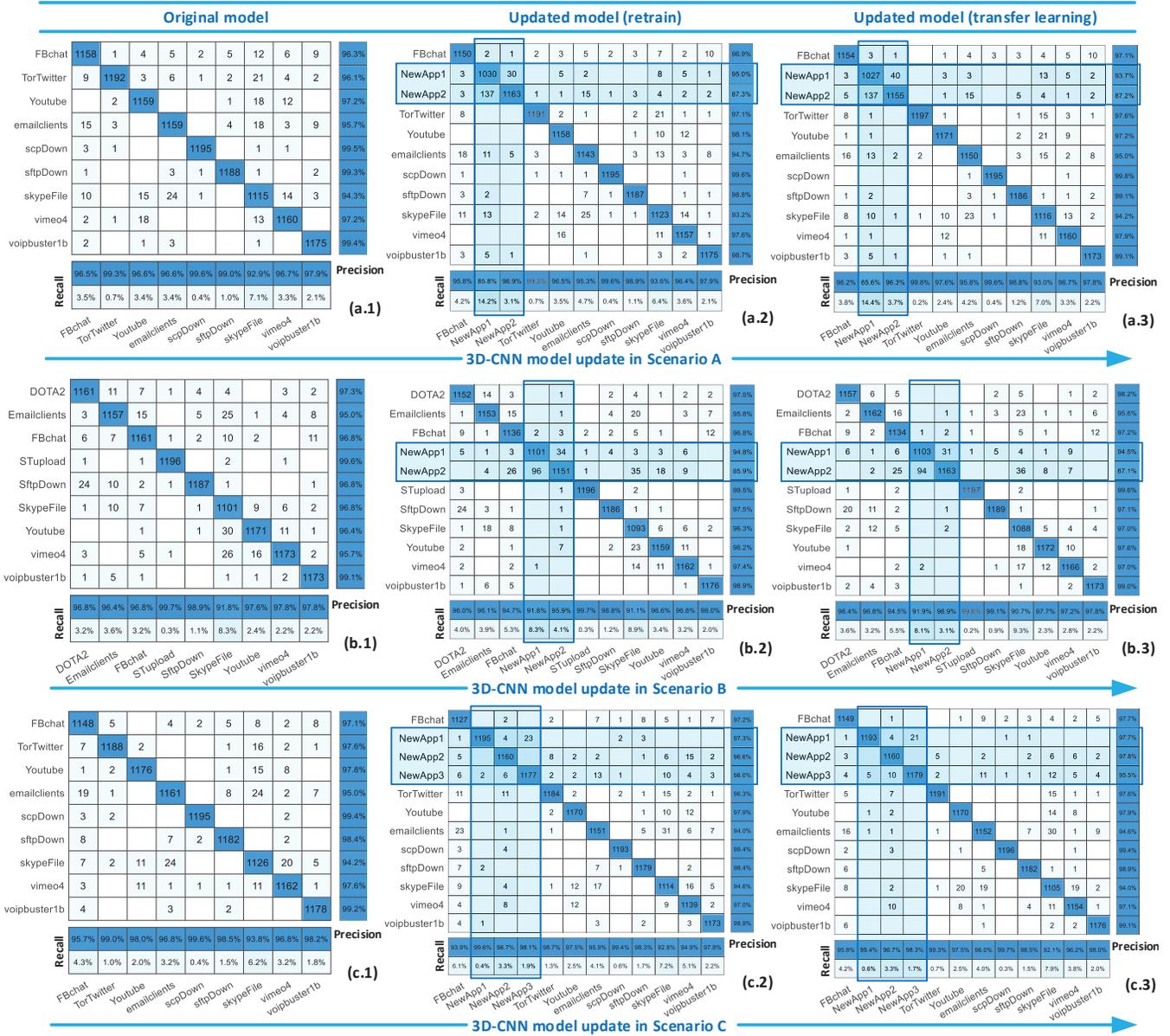}
\caption{Performance confusion matrices of original, updated (retain), and updated (transfer) 3-D CNN classifiers.}\label{12tu} 
\end{figure*}

We further analyze the 3D-CNN based classifier in the aspects of 1) the clustering performance when assigning labels to the unknown classes, and 2) the classification performance of both original and updated classifiers in all proposed scenarios. In the autonomous clustering process of the model update for 3D-CNN classifiers, we performed feature extraction by choosing the first 7 most significant components in PCA to preserve 95\% content of the original feature maps. K-means clustering algorithm is then applied to assign the labels for the corresponding unknown classes. The number of clusters is determined autonomously according to Alg.~\ref{alg2}. The clustering results for 3D-CNN based classification model in all scenarios are presented below in Fig.~\ref{fig:awesome_image1}. The calculated Recall, Precision and F score are given in Table~\ref{tb6}.

The clustering performance summarized in Fig.~\ref{fig:awesome_image1} and Table~\ref{tb6} indicates that the proposed scheme can cluster multiple unknown classes with the corresponding packets filtered by the discriminator. 



To further evaluate the performance of 3D-CNN classifiers, we compose a testing dataset with data portion $\mathcal{P}_3$ (1200 packets from each existing application) and data portion $\mathcal{P}_5$ (1200 packets from each unknown application). Fig.~\ref{12tu} demonstrates the confusion matrices of classification results performed by the original classifier and the updated classifier with or without transfer learning. Each confusion matrix provides classification Recall and Precision at the bottom and to the right side respectively, which are accuracy metrics that indicate the classification performance. The diagonal of each confusion matrix presents the amount of packets of each application that are correctly classified. In all Scenarios, the observation shows that all classifiers have a high overall classification accuracy. However, due to the involvement of new classes, the accuracy of both updated classifiers for each existing class is slightly lower than the original classifier. In either Scenario A or Scenario B, almost all new classes are classified properly. Note that the accuracy of classifying Google Map is relatively lower than others in updated classifiers. It is because of the similarity of packets between Google Map and another unknown application appeared in Scenario A. Moreover, in Scenario C, the classification performance of new classes is superb, which reaches an average of 97\%.

\section{Conclusion}\label{sec6}
Network traffic classification is the fundamental for accurate network measurement and efficient network management. To solidify those classifiers in an open-world assumption, we proposed an autonomous model updating framework that can filter the packets of unknown classes and cluster them to the corresponding classes. The filtered packets with their assigned labels as well as the packets of existing classes are combined to produce a new dataset to update the classifier. To evaluate the proposed framework, we used the packets captured in real life and the packets in an open dataset. Moreover, three scenarios were designed by mixing different classes in the dataset. The evaluation results demonstrated that our proposed autonomous model updating framework can update DL-based traffic classifiers with the capability of classification of the packets from unknown classes in active network.



\renewcommand\refname{Reference}
\bibliographystyle{IEEEtran}
\bibliography{IEEEfull,Reference}
	
\end{document}